# GanglionNet: Objectively Assess the Density and Distribution of Ganglion Cells With NABLA-N Network

Md Zahangir Alom, *Member*, IEEE, Raj P. Kapur, TJ Browen, and Vijayan K. Asari, *Senior Member, IEEE*

**Abstract**—Hirschsprung's disease (HD) is a birth defect which is diagnosed and managed by multiple medical specialties such as pediatric gastroenterology, surgery, radiology, and pathology. HD is characterized by absence of ganglion cells in the distal intestinal tract with a gradual normalization of ganglion cell numbers in adjacent upstream bowel, termed as the transition zone (TZ). Definitive surgical management to remove the abnormal bowel requires accurate assessment of ganglion cell density in histological sections from the TZ, which is difficult, time-consuming and prone to operator error. We present an automated method to detect and count immunostained ganglion cells using a new NABLA-N network-based deep learning (DL) approach, called "GanglionNet". The morphological image analysis methods are applied for refinement of the regions for counting of the cells and define ganglia regions (a set of ganglion cells) from the predicted masks. The proposed model is trained with single point annotated samples by the expert pathologist. The GanglionNet is tested on ten completely new High-Power Field (HPF) images with dimension of 2560×1920 pixels and the outputs are compared against the manual counting results by the expert pathologist. The proposed method shows a robust 97.49% detection accuracy for ganglion cells, when compared to counts by the expert pathologist, which demonstrates the robustness of GanglionNet. The proposed DL-based ganglion cell detection and counting method will simplify and standardize TZ diagnosis for HD patients.

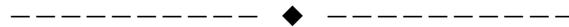

## 1 INTRODUCTION

NEURONS or ganglion cells are essential to cognition and most other bodily functions. Many human diseases result from congenital or acquired conditions that increase or reduce specific neuronal populations. These include Hirschsprung disease (HD or HSCR), a birth defect in which ganglion cells are absent in the distal rectum and a variable length of contiguous and more rostral bowel. Intestinal ganglion cells are normally located in small groups or ganglia, which are found in the myenteric (located in the muscular layers of the outer bowel wall) and submucosal (located in connective tissue in the inner half of the bowel wall) plexuses. In the aganglionic segment of HD, both plexuses are devoid of ganglion cells. Patients with HD usually present shortly after birth with signs of intestinal obstruction such as failure to pass stool, abdominal distension, and vomiting; less severe cases are diagnosed later in the childhood or rarely in adults. HD affects one-in-five thousand liveborn infants and can have lethal consequences. It is treated surgically by removing both the abnormally innervated distal bowel and attaching more proximal normoganglionic bowel to the anus. It is important to resect not only the distal aganglionic segment, but also the transition zone (TZ), a variable length of neuroanatomically abnormal intestine immediately upstream from the aganglionic segment.

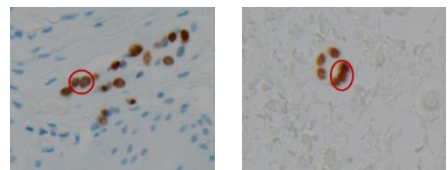

Fig. 1. H and N types of scans: sample for H type of scan on the left and N type of scan on the right. The brown regions within the images represent the target ganglion cells and a set of ganglion cells create ganglia. Red circles highlight contiguous nuclei that must be discriminated from one another.

Histopathological identification of TZ is used to guide surgery and post-operative management. A consistently observed microscopic feature of TZ is a gradual transition in myenteric ganglion cell density from complete absence to normal. In addition, in some patients, the proportion of large (increased number of ganglion cells per ganglion) submucosal ganglia is abnormally high (HD-associated intestinal neuronal dysplasia), although the functional consequence of this pattern of submucosal hyperganglionosis is controversial [1,2]. Diagnosis of these alterations is subjective and usually based on a pathologist's non-

- *Md Zahangir Alom is with the University of Dayton, Dayton, OH 45469, USA. E-mail: alomm1@udayton.edu.*
- *Raj P. Kapur. is with the Seattle Children's Hospital and University of Washignton School of Medicine, WA 98195, USA. E-mail: raj.kapur@seattlechildrens.org.*
- *T.J Bowen is with DeepLens, Columbus, OH 43212, USA. E-mail: tj@deeplens.ai*
- *Vijayan K. Asari is with the Department of Electrical and Computer Engineering, University of Dayton, Dayton, OH 45459, USA. E-mail: vasari1@udayton.edu.*

quantitative impression or, less commonly, formal neuronal counts. However, even the latter are influenced greatly by subjective factors, such that inter-observer concordance is unsatisfactory.

We propose to use Phox2B immunohistochemistry (ihc) and pathological image analysis method with deep learning to minimize operative bias from histological diagnosis. Phox2B is a transcription factor, which is specifically expressed in the nuclei of postnatal enteric neurons. Phox2B ihc is suitable for most clinical laboratories as it can be performed on formalin-fixed paraffin-embedded tissue sections with an automated ihc system and conventional peroxidase-based immunochemistry with (H) or without (N) hematoxylin counterstain (to facilitate resolution of background tissue architecture). Pathological image analysis procedures represent a "cutting-edge" in the field of diagnostic pathology as they promise to eliminate subjectivity, increase accuracy, and potentially reduce cost. Potential obstacles to obtain accurate neuronal counts from Phox2B-immunostained sections include discrimination between labeled and hematoxylin-stained unlabeled nuclei and variations in nuclear size and staining intensity, due in part to the nature of histology, overlap/contiguity between labeled nuclei, and the fact that only portions of individual nuclei will be present in any given tissue section.

In this study, we introduce and test an automated counting system for ganglion cells and ganglia. The model is not only able to detect and discriminate the Phox2B-labeled nuclei of individual ganglion cells, but also accurately count the number of cells and define ganglia in high power field (HPF, 200×) images. In most of the scans, the nuclei of individual ganglion cells are isolated but some cases they overlap and appear connected (red circles in Fig. 1). This problem makes ganglion cell counting especially challenging. The statements of novelty and impact of this paper are as follows:

- For the first time, the Deep Learning (DL)-based enteric ganglion cells detection and counting system is proposed for the assessment of ganglion cell density in histological sections from the TZ in HD.
- NABLA-N Net based deep model is proposed and applied for ganglion cells detection tasks and morphological image analysis methods are applied for refinement of the regions for counting of the cells and define ganglia regions (a set of ganglion cells).
- The training is done with single point annotated samples which are annotated by the expert pathologist.
- Testing is performed on ten completely new high-power fields (HPF, 200x images; 2560×1920 pixels) and the performance is compared against manual analysis done by the expert pathologist.
- NABLA-N Net shows 97.49% ganglion cells detection in terms of F1-score compared to the results in manual counting which demonstrates the robustness of GanglionNet.

The rest of the paper is organized as follows: related works are discussed in section 2. Section 3 explains the ganglion cells detection system and models. Databases, results and discussion are given in section 4. Finally, the conclusion is given in section 5.

## 2 RELATED WORKS

Failure to surgically resect the TZ can result in persistent obstructive symptoms after HD surgery and may require a repeat operation. Therefore, recent studies have emphasized diagnostic histological criteria for TZ, which are based in large part on the density and distribution of enteric ganglion cells [1,2]. However, it is generally acknowledged that conventional assessment of ganglion cells is subjective and operator-dependent, such that mild or moderate alterations of ganglion cell number are likely to be under-recognized. Therefore, we propose a deep learning-based ganglion cells and ganglia detection and counting system for defining the cell density in TZ. Computerized image analysis has been used to discriminate immuno-labeled nuclei or other cellular components in histological sections for a variety of research and clinical contexts [3]. Indeed, in the context of HD, a machine learning approach has been shown to differentiate between aganglionic and ganglionic tissue sections, which were immunostained with a variety of cytoplasmic neural markers [4]. However, the latter method did not allow for quantification of ganglion cells and was not applicable to delineation of the TZ. To the best of our knowledge, no published study has targeted automated enteric ganglion cell and ganglia region detection, especially in relation to TZ diagnosis and the management of patients with HD. In addition, the importance and a role of computerized morphometric analysis for diagnosis of HD was clearly demonstrated in 2010 [5].

In 2019, a very closely related work published a three-stage identification method, called "MapDe", and that was applied for peak detection of cells. MapDe was proposed based on the ability to learn and regress the probability of each pixel belonging to the center of the cells followed by the detection of local maxima. MapDe was evaluated on two different datasets for lung adenocarcinomas and achieved better detection of 79.06% in terms of F1-score [6]. Another peak detection method has been trained on artificial labels and performed peak detection [7]. Recently, the Cell Count RegularizeD Convolutional Neural Network (ConCORDe-Net) was developed for cell detection and counting in multiplex immunohistochemistry images. The ConCORDe-Net combines a convolutional dice overlap and a new cell counting loss function for optimization of cells detection [8]. An active cell appearance model (ACAM) has been proposed which can measure the statistical distribution of shape and intensity of the cells. ACAM, when used with a guide Conditional Generative Adversarial Model (CGAN) and referred to as "AGAN", was evaluated for cells analysis in adaptive optics retinal imaging and achieved substantially improved results [9].

In all these digital pathology applications, overlap or closely located cells with weak boundaries make cell counting difficult. In addition, the size of the kernel has a big impact of cell counting problem as highlighted by a recent study, which stated that the size of the kernel should not outfit with a range of size and shape [10,11,12]. All these related works tried to detect and count the different cells in digital pathology application, however, in this paper, we propose an end-to-end ganglion cells detection and counting for the very first time and shows promising results compared to expert pathologist.

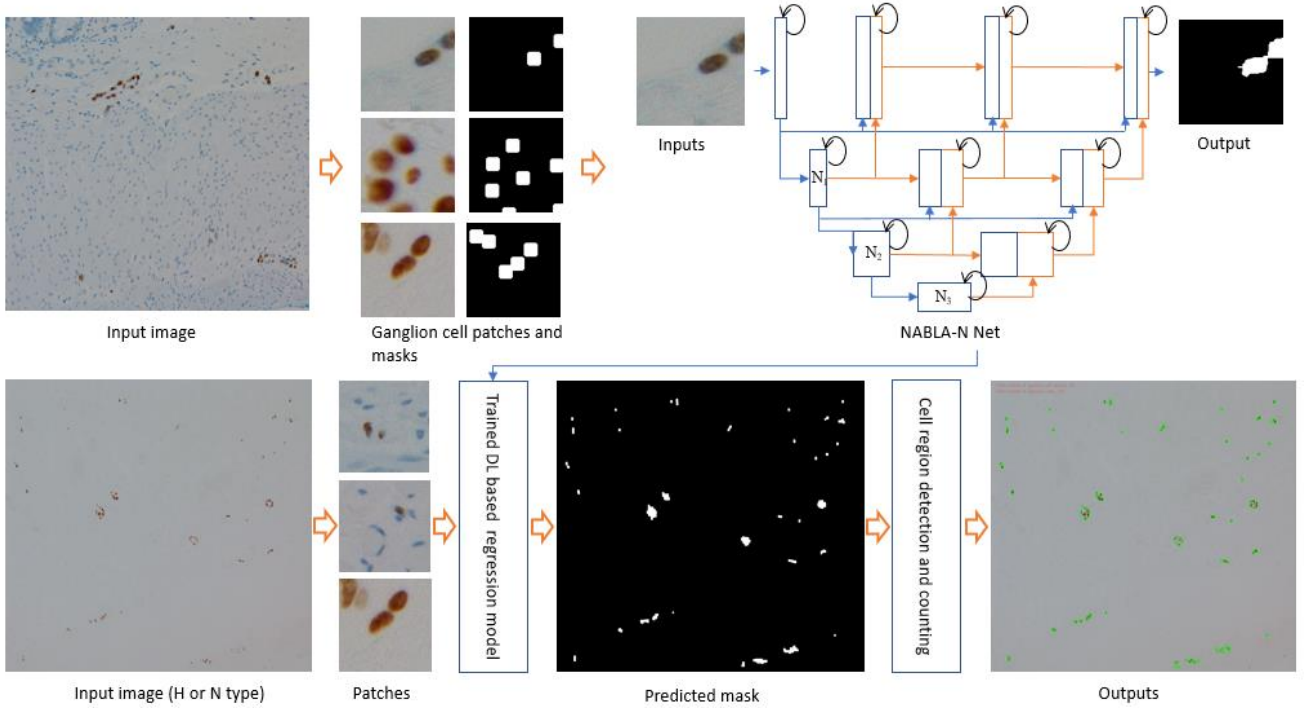

Fig. 2. End-to-end training and testing method for ganglion cell detection: top row shows the training process including input image, patching, and training with NABLA-N Net and the bottom row shows the testing pipeline which includes input image, patch extraction, testing and results

## 3 METHODOLOGY

**Digital images from Phox2B-immunostained tissue sections:** Full-thickness samples of colon wall from the TZ of a patient with HD were fixed in formalin, embedded in paraffin, and used to prepare 4-5 μm-thick histological sections. Phox2B (rabbit antibody, AbCam Ab183741; 1:1000 dilution) ihc was performed with a Ventana Benchmark immunostainer (standard conditioning, 32 min primary incubation at 37ºC). Digital images at 200x magnification were captured with a Nikon photomicroscopy system (Eclipse microscope, Sight DS camera, and Elements BR software). The digital images were annotated by one of the authors (RPK), a board-certified pediatric pathologist with HD expertise.

**Overview:** We propose an enteric ganglion cell and ganglia region detection and counting approach with NABLA-N net model for the very first time. The entire processing pipeline diagram is shown in Fig. 2. The top row shows the training process and the bottom row shows the testing pipeline. For the training process, the H- and N-type samples are collected with single point annotations where one single pixel represents the entire ganglion cell. Hence, we have generated Gaussian density surface with respect to the center point of the cells and dilated the HPF image. The Gaussian distribution is represented with G in the following equation and the thresholding is performed with respect to the values of 0.5 to generate the segmentation task.

$$M(i,j) = \begin{array}{ll} 1 & if \; dilated((G > 0.5), k) \\ 0 & otherwise \end{array}$$

The masks are then dilated with kernels with a kernel size of k =7, 9, 11, and 13. Then, 128×128 pixels patches and masks are extracted form 2560×1920 pixels input image and corresponding masks with respect to the availability of ganglion cells. The bottom part of Fig. 1 shows the testing pipelines for both H and N types tissue. For testing, the entire HPF images are used as input to the system, the non-overlapping patches are extracted from the input HPFs and then generated the outputs with trained NABLA-N network. At the end, the small patches are merged to produce the HPF level mask. The cell counting is performed on HPFs output masks.

**NABLA-N net model:** The NABLA-N net is an improved deep learning architecture for medical image segmentation which was proposed in 2019 [13]. According to the deep learning architecture for segmentation task, both lower and higher level of features are equally important to ensure better training and define the proper boundary for the segmented regions. The NABLA-N net was designed considering this fundamental phenomenon for segmentation tasks and shown higher accuracy. Here, N refers the number of encoding feature representation layers (including bottleneck layer) which are considered for decoding the features [13]. The U-Net type models utilize the features from all levels of representation (low to high level features) in encoding units and concatenate with the features with decoding units which are generated from only the bottleneck layer [14]. However, the NABLA-N network decodes the features from N different encoded layers and fuse between them. As a result, it ensures better feature representation.

The conceptual diagram of NABLA-3 ($N_{1-3}$) net is shown at the end of the top row in Fig. 2, where $N_1$ represents the low-level features (edges, color, texture etc.) $N_2$ layer represents the middle level features including the contextual features related to the part of the images whereas $N_3$ layer represents the high-level features of the entire input images. As the features are decoded from different levels of representation and fused between them, therefore, the proposed model ensures better feature extraction and representation compared to the U-Net type models [14, 15]. There are several advantages for this proposed architecture: first, this architecture ensures feature reusability which is inspired by the very popular DenseNet model [16]. Second, since, instead of using the straightforward convolutional layers, the recurrent convolutional layers (RCL) are used in each convolutional unit in both encoding and decoding units [17,18]. As a result, the recurrent operations help to extract and preserve the missing features from the target regions compared to straightforward convolutional layers. Third, this capability can significantly reduce the network parameters while it shows better performance for segmentation tasks. Thus, the model becomes lighter and faster.

**NABLA-N net architecture:** We evaluated the NABLA-N net models with 13 layers including input and output layers. The entire model architecture is as follows: 3→16×(3×3) →32×(3×3) →64×(3×3) →128×(3×3) →256×(3×3) →512×(3×3) →256×(3×3) →128×(3×3) →64×(3×3) →32×(3×3) →16×(3×3) →1. The number of feature maps and kernel sizes can be represented with general notation of $F_N$× (M×N) where $F_N$ represents the number of feature maps and (M×N) represents the kernel size which is (3×3) kernels used in each layer except the last layer. The recurrent convolutional operations are performed with respect to the time steps t = 2 which means one forward convolutional layer is used followed by two recurrent layers. We have used Rectified Linear Unit (ReLU) activation function in each layer [19]. At the end, a 1×1 convolutional layer is used to reduce the dimension of the feature maps from 128×128×16 dimension to 128×128×1, then a sigmoid activation function is applied. In addition, we have utilized the up-sampled feature maps from 3 different encoding layers including the bottleneck layer, therefore, we named the model NABLA-3 network. The NABLA-3 model utilizes totally 18.98 Million (M) network parameters. The network is initialized with He initialization method [20].

**Ganglion cells counting:** Since the clinical and pathological assessment is done on the HPF level, we here present the ganglion cell detection and counting on HPF images. In the testing phase, non-overlapping patches (128×128 pixels) are extracted from input images of size 2560×1920 pixels which is shown in Fig. 2. The model segments both the ganglion cells and ganglia regions (a group of the ganglion cells) accurately. Eventually, all the patches are merged together to produce the final output for the entire image. Then, the morphological operations are performed to refine the output masks generated by the NABLA-3 network, since, it is very hard to count the number of ganglion cells from the connected cells regions. To resolve this problem, the trained model is tested on the training HPFs to find the average number of pixels for individual ganglion cell region for H and N type of tissue. Then, we utilized this average number of pixels to define independent ganglion cells and ganglion cell regions. The experimental results show very accurate number of ganglion cell counting from input HPF images.

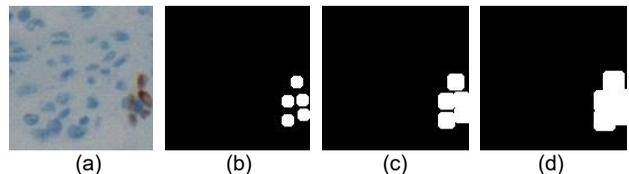

Fig. 3. Input patches and masks with respect to different kernel sizes: (a) input, (b) dilation kernels 5×5, (c) dilation kernels 9×9, and (d) dilation kernels 13×13.

## 4 EXPERIMENTS AND RESULTS

We implemented the ganglion cell detection method named GanglionNet with TensorFlow deep learning framework on four NVIDIA GTX2080 Ti single GPUs. For testing, the end-to-end system was run on the single GPU system. The proposed NABLA-N network-based cell detection model was evaluated on our own dataset. The samples were annotated with single point annotation method (means one single point represent the entire cell) by expert pathologist.

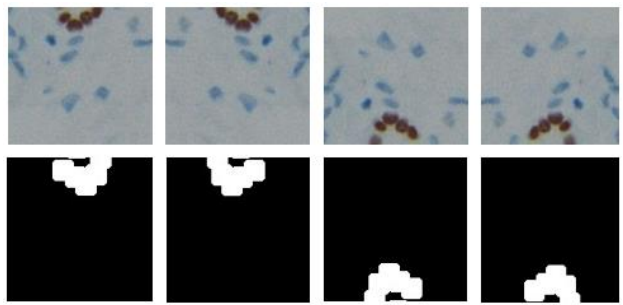

Fig. 4. Randomly selected augmented patches and corresponding masks from the training samples.

Therefore, this annotation method makes this problem more difficult to detect and count ganglion cells. The database details and the experimental results are discussed in the following sections.

## 4.1 Database

For training and validation this proposed system, a total of 12 High Power Field (HPF) samples were collected in the first phase and the size of the image is 2560×1920 pixels. The samples were collected with different magnification factors that include 100× and 200×. Ten different samples (HPFs) were randomly selected for training and remaining two samples were used for testing. We evaluated the proposed model with fivefold validation methods. Most of the pixels of the input patches represent background as shown in Fig. 3. As a result, it creates a class imbalance problem for this task. To resolve this class imbalance problem, we have extracted patches from only regions where the target cells present. Total 514 patches were extracted from ten training samples and the size of the patch is 128×128 pixels. The input patches and masks with respect to different size of the dilation kernels are shown in Fig. 3(b) to 3(d). It can be clearly seen from Fig. 3 that the cells are merged with each other with respect to the size of the dilation kernels. For selecting the optimum kernel size for training the NABLA-3 network model, we used the masks generated with 5×5, 9×9, and 13×13 sizes of the dilation kernels. To increase the number of samples, data augmentation operations (such as horizontal and vertical flipping) are performed. Total 2056 patches are used for training and validation of the model. The randomly selected augmented samples are shown in Fig. 4. We have investigated the performance of the model for all kinds of segmentation masks. The proposed model shows poor training and validation accuracy for both masks generated with 5×5, 9×9 kernels, however the model does not converge properly due to insufficient reference feature representation. We have found that the model learned and converged faster for the masks generated with the kernel size of 13×13 due to having enough reference features to be learned. Thus, 13×13 kernels-based model is used for final evaluation. Furthermore, the size of the dilation kernels plays a big role to define the individual cell during the testing phase as well.

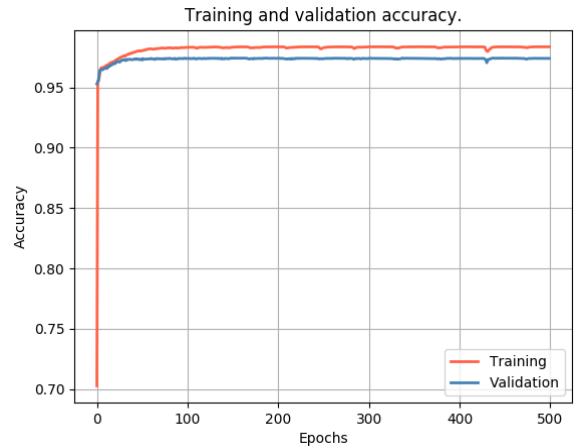

Fig. 5. Training and validation accuracy for ganglion cell detection task. The orange color shows the training dice coefficient (DC) and blue color represent the validation accuracy.

**Training method:** As the number of labeled samples is low, the GanglionNet is trained for 500 epochs. The network is trained with Adam optimizer with a learning rate of $3 \times e^{-4}$, binary cross entropy loss, and batch size of 32. This experiment shows around 98.5% training accuracy and around 97% validation accuracy for the ganglion cell detection task. Furthermore, since we did not have adequate samples for training the GanglionNet, therefore, we have applied data augmentation methods without changing the actual image properties and the examples augmented samples and masks are shown in Fig. 4. We have trained the model for 500 epochs. The training and validation accuracy are shown in Fig. 5. The experiment shows around 98.5 percentage of training accuracy and around 97 percentage of validation accuracy for ganglion cell

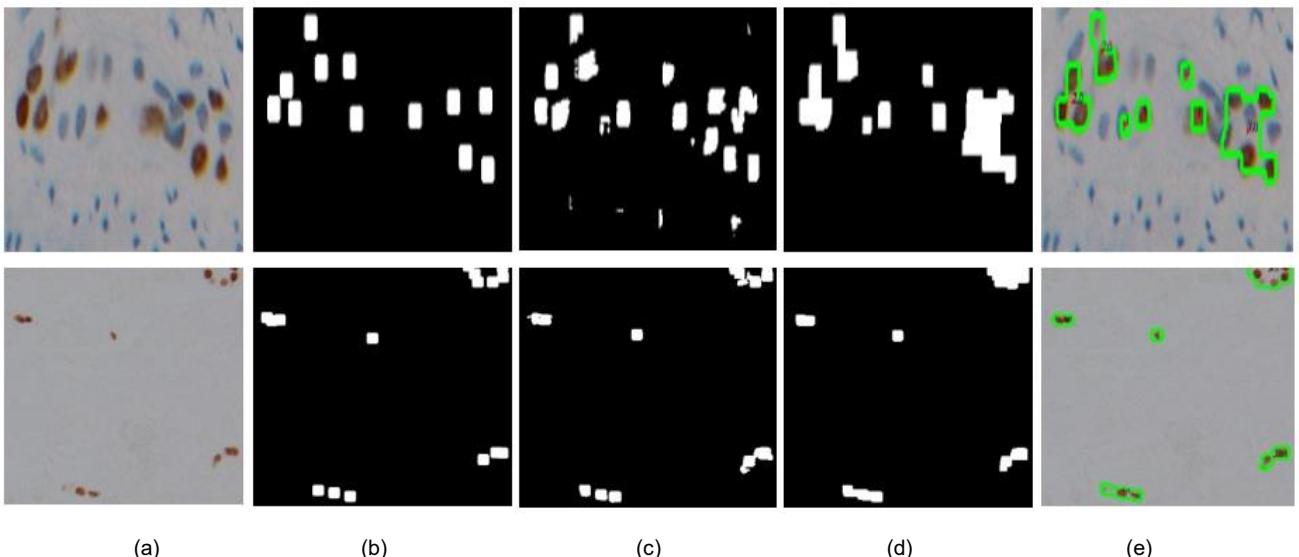

(a) (b) (c) (d) (e)

Fig. 6. Experimental output for H-type scan (magnification factor 200x) and N-type scan (magnification factor 100×): the first row shows results for H type input image; second row represent the results for N type of scan. (a) Inputs (b) Ground Truth (GT) (c) Model outputs (d) Outputs after performing morphological operations, and (e) shows the final outputs with ganglion cell regions, contour of the region, and total cells count.

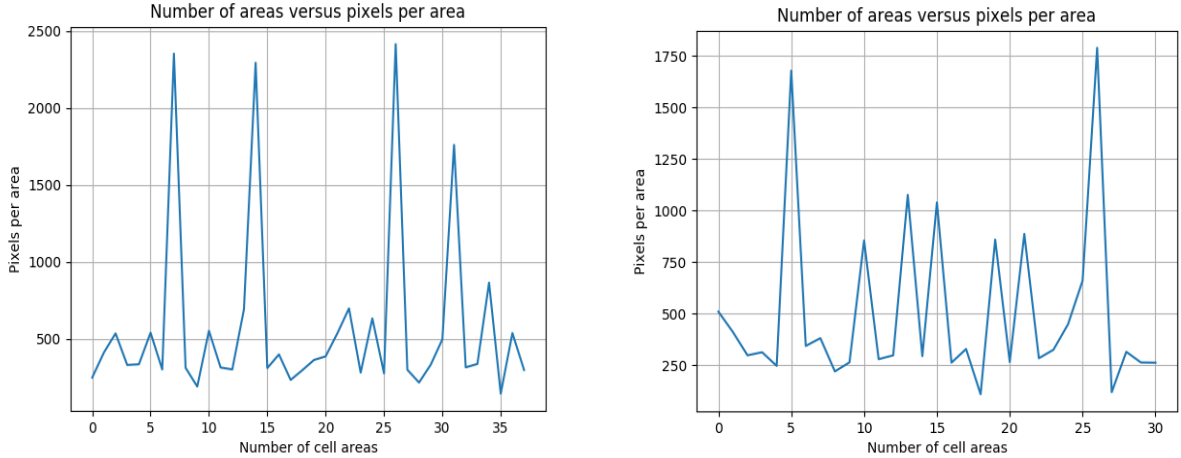

Fig. 7. The size of the ganglion cell areas in term of number of pixels: left plot shows the average areas of thirty-seven ganglion cell areas for H-type scan and right plot shows the average areas for N-type scans for thirty areas.

detection task.

TABLE 1
QUANTITATIVE RESULTS FOR H AND N TYPES TISSUE SCANS

| Input ID | Number of cells | |
|---|---|---|
|  | Manual counting | Predicted |
| H_field_3_20x | 43 | 41 |
| N_field_1_10x | 104 | 104 |

### 4.2 Results

After training the model successfully, the GanglionNet is tested on twelve completely separate H and N type of HPFs (2560×1920 pixels). We performed thresholding operation with respect to the value of 0.5 on the model outputs. Fig. 6 shows the outputs for part of H and N types of samples in the first and second rows respectively. In Fig. 3, second column shows the ground truth (GT), the third column shows the model outputs and forth column shows outputs after performing morphological operations with kernel size 13×13. The final detected regions with contour and number of cells of the regions are shown in the fifth column. This model is also able to define the ganglia regions (a group of ganglion cells) successfully. To define the number of individual cells and total count of the cells, we have calculated the average number of the pixels for the individual ganglion cells. The number of pixels per cell's region of randomly selected H and N type scans are shown in Fig. 7. Finally, we have empirically calculated average number of pixels for individual cell regions for H and N types of scans and are around 350 pixels and 270 pixels respectively. These pixel averages are used to define individual ganglion cells and ganglia regions. Therefore, we have designed an automatic counting method with respect to the scan types.

The quantitative analysis for the entire H and N types input images are shown in TABLE I. The model is evaluated with two samples with 200× and 100×, however, the model shows very accurate number of ganglion cells compared to manually counting by the expert pathologist. For H type scan, the total number of ganglion cells are 43 whereas the model provides 41 as outputs. On the other hand, the manual cell counting for N type input sample is 104 and model shows 104. Thus, our proposed model is very accurate and robust for counting ganglion cells compared to manual counting by the expert pathologist.

### 4.2 Comparison

After training and evaluating the model successfully, the performance of the proposed detection and counting system is evaluated on 10 completely new samples provided by the pathologist. We have appraised the performance of the proposed model with different sizes of the kernels, 11×11 ($K_{11}$), and 13×13 ($K_{13}$), and comparison against manual counting by the expert pathologist are shown in the Table 2. For 11×11 kernels, we have calculated average number of pixels for H and N type samples as 320 and 250 pixels respectively for individual cells. The cell counting significantly vary because of the non-target regions (a set of pixels) inside the ganglion regions which can be clearly observed from Figs 6. Hence, for H type input sample, if the total number of pixels is higher than 900 pixels then floor(toal_pixel_per_resion/average_pixels_per_cell_H) – 1 is considered for cell counting. If the total number of pixels of the ganglia regions is more than 1250 pixels then floor(toal_pixel_per_resion/average_pixels_per_cell_H) – 2. On the other hand, since the ganglion cells in N type inputs are smaller in size compared to H type and unwanted pixels regions are less inside the ganglion cell regions. Thus, we directly applied floor(toal_pixel_per_resion/ average_pixels_per_cell_N) for counting number of ganglion cells for N-type inputs. All these values are defined empirically.

The results are then compared against manual counting by the expert pathologist. The model demonstrates extremely good concordance between "ground truth" manual counts performed by the pathologist for ten different samples shown in Table 1. We have achieved 91.25% and 97.49% testing accuracy for dilation kernels of 11×11, and 13×13 respectively. The qualitative results for H&N types of input samples and outputs are shown in Fig. 8. First and second rows represents outputs for H type samples and third and fourth rows show results for N type input samples. First column shows the input images with manual counting by pathologist, second column shows the outputs from model, third column shows refined outputs after performing morphological operations, and fourth column shows final outputs with contours and number of cells within the regions. The GanglionNet provides very accurate ganglion cells and ganglia region detection with respect to manual counting by the expert pathologist.

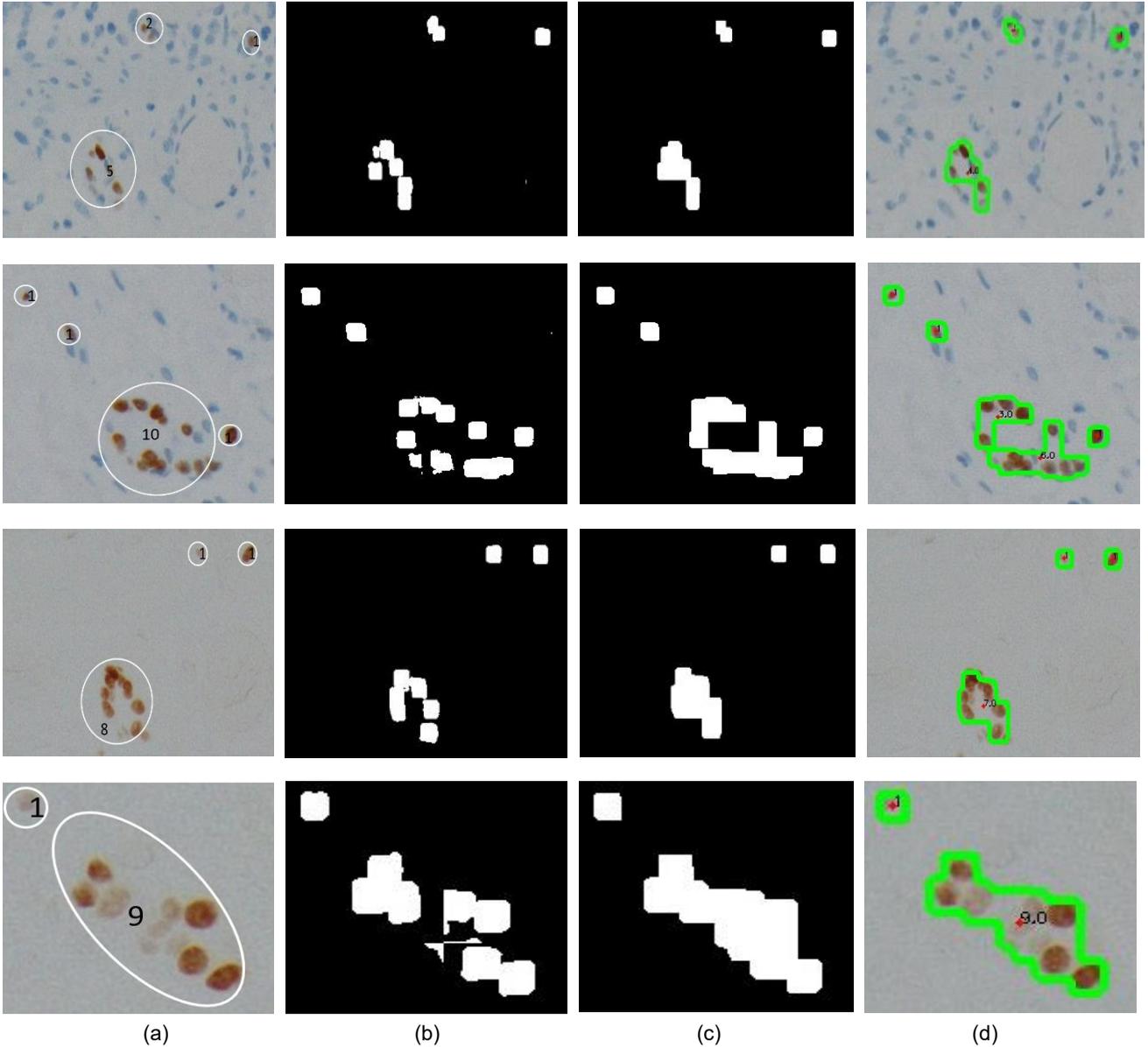

Fig. 8. Results and comparison against manual counting by the expert pathologist: (a) manual counting by pathologist (b) detected results with NABLA-3 model, (c) outputs after refinement with morphological operations, and (d) final outputs.

**Computational time:** The computational time vary with respect to the configuration of the computing system. According to our system setup, the model takes around 75 seconds per epoch for training which is around 10.41 hours in total. The end-to-end testing is performed on input images with size of 2560×1920 pixels. The GanglionNet takes only 11.44 seconds for analyzing the entire input image, counting, and for the representation of the results. Thus, we believe the proposed model is fast enough to deploy the end-to-end system for clinical practice.

TABLE 2
NUMBER OF GANGLION CELL CONTING WITH RESPECT TO DIFFERENT SIZE OF KERNELS ($K_{11}$, AND $K_{13}$) AND COMPARISON AGAINS THE MANUAL COUNTING ($M_C$)

|  | Number of ganglion cells |  |  |  |  |  |  |  |  |  |
|---|---|---|---|---|---|---|---|---|---|---|
| IDs | 1 | 2 | 3 | 4 | 5 | 6 | 7 | 8 | 9 | 10 |
| $K_{11}$ | 98 | 57 | 76 | 52 | 46 | 57 | 40 | 40 | 74 | 42 |
| $K_{13}$ | 90 | 65 | 71 | 54 | 45 | 60 | 44 | 41 | 75 | 41 |
| $M_c$ | 103 | 55 | 77 | 53 | 48 | 58 | 38 | 40 | 74 | 43 |

**Advantages:** We proposed an end-to-end scan invariant ganglion cell detection and counting system from H- and N-type of samples. The proposed system is scale invariant which works for both scans with 100× and 200× magnification factors. Both the quantitative and qualitative results demonstrate that the system is robust enough to detect the ganglion cells with very high accuracy. However, cell counting is a very challenging task which depends on size, scale, and density of the cells as well as input images which still need to be improved. **Limitation**: cell counting is a very challenging task which depends on size, scale, and density of the cells as well as input images. Although the proposed model provides accurate cells detection results, however, cell counting method does not provide accurate cell counting in some exceptional cases. There is a lot of scope to improve this cell counting part to make this system more robust and accurate.

## 5 CONCLUSION AND FUTURE WORKS

In this paper, we presented an DL-based ganglion cell detection and region-based cell counting method. We evaluated our proposed model with a set of experiments evaluated and annotated by an expert pathologist. The experimental results demonstrate the effectiveness of our method for ganglion cells, group of ganglion cells detection and counting tasks. The proposed method shows 97.49% accurate cell counting compared to manual counting by the expert pathologist. In future, we would like to train the model with more samples to ensure better performance, quantify ganglion cells at different points along the TZ from many HD patients, correlate our automated results with other indices of TZ histology, and experiment with hematoxylin and eosin-stained samples.

## ACKNOWLEDGMENT

Special thanks to DeepLens for funding and Dr. Kapur for annotating and providing this dataset.